# High-precision luminescence cryothermometry strategy by using hyperfine structure

Marina N. Popova[1]*, Mosab Diab[1,2], Boris Z. Malkin[3]

[1] *Institute of Spectroscopy, Russian Academy of Sciences, Troitsk, Moscow 108840, Russia*

[2] *Moscow Institute of Physics and Technology (National Research University), Dolgoprudnyi, 141700, Russia*

[3] *Kazan Federal University, Kazan, 420008, Russia*

**Abstract**

A novel, to the best of our knowledge, ultralow-temperature luminescence thermometry strategy is proposed, based on a measurement of relative intensities of hyperfine components in the spectra of $Ho^{3+}$ ions doped into a crystal. A $^7LiYF_4$:$Ho^{3+}$ crystal is chosen as an example. First, we show that temperatures in the range 10 – 35 K can be measured using the Boltzmann behavior of the populations of crystal-field levels separated by an energy interval of 23 cm$^{-1}$. Then we select the 6089 cm$^{-1}$ line of the holmium $^5I_5 \rightarrow {}^5I_7$ transition, which has a well-resolved hyperfine structure and falls within the transparency window of optical fibers (telecommunication S band), to demonstrate the possibility of measuring temperatures below 3 K. The temperature $T$ is determined by a least-squares fit to the measured intensities of all eight hyperfine components using the dependence $I(v) = I_1 \exp(-bv)$, where $I_1$ and $b = av + v/kT$ are fitting parameters and $a$ accounts for intensity variations due to mixing of wave functions of different crystal-field levels by the hyperfine interaction. In this method, the absolute and relative thermal sensitivities grow at $T$ approaching zero as $1/T^2$ and $1/T$, respectively. We theoretically considered the intensity distributions within hyperfine manifolds and compared the results with experimental data. Application of the method to experimentally measured relative intensities of hyperfine components of the 6089 cm$^{-1}$ PL line yielded $T = 3.7 \pm 0.2$ K. For a temperature of 1 K, an order of magnitude better accuracy is expected.

* corresponding author, e-mail: popova@isan.troitsk.ru

# I. INTRODUCTION

Luminescent thermometry is a remote method of temperature measurement based on the temperature dependence of various luminescence properties such as intensity, polarization, decay time, lines' spectral position and width [1-7]. Inorganic and organic micro- and nanocrystals containing rare-earths (RE) or transition-metal ions, diamonds with color centers, quantum dots, and hybrid organic-inorganic materials are used as luminescence sources. The method has developed rapidly over the last two decades, especially for the biologically important temperature range. Besides, the luminescent thermometry is used for controlling chemical reactions, detecting phase transitions, internal temperature monitoring of Li-batteries, in catalysis, etc [3-5]. In connection with these applications, the temperature range 80 – 800 K is well mastered. At the same time, areas such as superconducting magnets, aerospace research, synchrotron crystallographic measurements, and modern quantum technologies require the measurement of lower temperatures, down to Kelvin and even lower.

In the range of temperatures lower than ~ 15 K, positions, widths, and lifetimes of energy levels of RE or transition-metal ions in crystals practically do not change. Here, Boltzmann luminescence ratiometric thermometry is the most suitable method. It is based on measuring the lines' intensity ratio $LIR(T) = I_i / I_j$ of two luminescent lines originating from two thermally coupled energy levels, $i$ and $j$ of a luminescent center, with degeneracies $g_i$ and $g_j$, respectively, separated by an energy interval $E_i - E_j = \Delta E$ [1-3]. Equilibrium populations of the levels obey the Boltzmann distribution:

$$\frac{n_i}{n_j} = \frac{g_i}{g_j} \frac{e^{-E_i/kT}}{e^{-E_j/kT}} = \frac{g_i}{g_j} e^{-\Delta E/kT}, \tag{1}$$

and thus

$$LIR(T) = \frac{I_i(T)}{I_j(T)} = \frac{W_i n_i(T)}{W_j n_j(T)} = C\, e^{(-\Delta E/kT)}, \tag{2}$$

where $W_i/W_j$ is the temperature-independent ratio of radiative transition probabilities, $k$ is the Boltzmann constant.

The absolute thermal sensitivity in this case

$$S_a(T) = \frac{d[LIR(T)]}{dT} = C \frac{\Delta E}{kT^2} e^{-\Delta E/kT} \tag{3}$$

has a maximum at $T_m = \Delta E/2k$. The lower is the measured temperature the smaller $\Delta E$ should be chosen. In particular, for the temperatures of the order of 10 K, 1 K, 0.1 K an adequate $\Delta E$ is 14 cm$^{-1}$, 1.4 cm$^{-1}$, 0.14 cm$^{-1}$, respectively. $\Delta E$ in the range from several to several tenths of wave numbers can be found as energy intervals between crystal-field (CF) levels of RE ions doped into a crystalline matrix. A sensitive low-temperature luminescence thermometry has been demonstrated utilizing

Boltzmann behavior of populations of the CF levels of $K_2YF_5:Er^{3+}$ [8], $LiYF_4:Er^{3+}$ [9], and $CaNb_2O_6:Pr^{3+}$ [10].

Here, we propose to use the hyperfine structure (HFS) of the $Ho^{3+}$ spectral lines in a crystal for a measurement of ultralow temperatures. A high precision of the measurement can be achieved by using all components of the HFS. We realize this strategy on the example of low-temperature photoluminescence (PL) of $^7LiYF_4:Ho^{3+}$. The hyperfine structure of energy levels originates from the interaction of electrons with the magnetic and electric quadrupole moments of the nucleus, the magnetic hyperfine interaction being the dominant one. The only holmium isotope $^{165}Ho$ has a large nuclear spin moment $I = 7/2$, which combined with a large hyperfine constant provides the most extended HFS consisting of eight components (in the case of the $Ho^{3+}$ ion residing in a sufficiently symmetric position of a crystal).

To investigate the applicability of HFS for the measurement of low temperatures, we selected a $^7LiYF_4:Ho^{3+}$ crystal. $LiYF_4$ crystals have excellent mechanical and thermal properties, are chemically stable, and possess a short phonon spectrum. Doped with RE ions, they have long been used as master single-mode oscillators for high-power lasers, as media for multifrequency and upconversion lasers, as upconverters to improve the power conversion efficiency of solar cells, and as nanoluminophors [11-16]. More recent applications include laser cooling [17-19] and temperature sensing [20-27]. In particular, $LiYF_4:Yb^{3+}/Er^{3+}$, $LiYF_4:Yb^{3+}/Tm^{3+}$, $LiYF_4:Yb^{3+}/Nd^{3+}$, and $LiYF_4:Pr^{3+}$ were used to measure temperatures about 300 K and above, while the interval 78-300 K was mastered with the help of $LiYF_4:Yb^{3+}$ [22]. Recently, a potential for measuring temperatures in the region 3 – 30 K was demonstrated for $LiYF_4:Er^{3+}$, based on relative intensities of the luminescence lines starting from two close in energy crystal-field levels [9].

HFS in the absorption [28-32] and luminescence [33] spectra of $LiYF_4:Ho^{3+}$ has been well studied. Monoisotopic $^7LiYF_4:Ho^{3+}$ crystal is preferred for sensing purposes, because its spectra are free from additional fine structure within HFS, caused by isotopic disorder in the lithium sublattice [32]. While the energy structure of the observed spectra was reproduced well by modeling based on crystal-field calculations [28,29,31], the intensity distributions within hyperfine manifolds, influenced by intermixing of wave functions of close in energy CF levels, were not studied. The aim of the present work was to fill this gap and to develop a method of cryothermometry based on HFS intensity measurements.

The paper is organized as follows. In Section II, we describe the sample used and details of optical measurements. Section III reports on the temperature-dependent luminescence spectra of $^7LiYF_4:Ho^{3+}$. In Section IV, the intensity distributions within hyperfine manifolds are considered theoretically and the results are compared with experimental data. In Section V, an approach is proposed of determining the temperature from the measured intensity distribution of hyperfine components. The paper ends with Conclusions.

## II. EXPERIMENTAL

A single crystal of $^7$LiYF$_4$:Ho$^{3+}$ (0.1 at. %) was grown in the Optical State Institute in St. Petersburg by the Stockbarger method using a mixture of monoisotopic $^7$LiF, YF$_3$, and HoF$_3$ fiuorides [32]. Photoluminescence spectra at different temperatures were recorded on a Bruker IFS 125 HR Fourier spectrometer with a resolution of up to 0.01 cm$^{-1}$. A luminescent attachment of our own design, CaF$_2$ beam splitter in the Fourier spectrometer, and an InGaAs detector with a high gain were used for recording the spectra (for more details on experimental setup see [33]). The crystal was cooled in a Sumitomo RP-082E2S closed helium cycle cryostat in steps of 0.5 K in the range of temperatures 3 K → 45 K. To excite luminescence, a multimode diode laser with a temperature-tunable wavelength 635 ± 10 nm was used. To reduce the thermal load on the sample, additional measures were taken, namely, filters were used to attenuate the excitation radiation, a double polished cold screen with small holes for radiation input and output was installed. The sample was glued with silver paste to the copper finger of the cryostat and additionally tightly wrapped with a sheet of metallic indium. Temperature was measured using a calibrated LakeShore DT-670 diode temperature sensor mounted in close proximity to the sample, controlled and recorded by a LakeShore Model 335 temperature PID controller. Temperature control was provided with an accuracy of ±0.05 K.

## III. EXPERIMENTAL RESULTS

Figure 1 shows the luminescence spectra of $^7$LiYF$_4$:Ho$^{3+}$ in the region of the $^5F_5 \rightarrow {}^5I_6$ and $^5I_5 \rightarrow {}^5I_7$ intermultiplet transitions. These transitions are of particular interest as they fall into the transparency windows of optical fibers (the U and S telecom bands, respectively). The luminescence lines are designated by the initial (capital Latin letters) and final (numbers) CF levels of the transition, where numeration starts from the lowest-energy level in the corresponding CF multiplet.

In the sheelite structure of LiYF$_4$, Ho$^{3+}$ ions substitute for Y$^{3+}$ ions in the $S_4$ symmetry positions. The wave functions of the holmium electronic CF levels transform according to the one-dimensional $\Gamma_1$ and $\Gamma_2$ and two-dimensional $\Gamma_{34}$ irreducible representations (IRs). Tables S1 and S2 within the Supplemental Material [34] clarify the lines' identification in Fig.1 but Table S3 highlights the nature of the observed transitions (electric- or magnetic-dipole). The $\Gamma_{34}$ CF levels are split by the magnetic hyperfine interaction into eight equidistant doubly degenerate electron-nuclear sublevels. The states $|\Gamma_3, m\rangle$ and $|\Gamma_4, -m\rangle$ have the same energy (here, $m$ is the component of nuclear moment $\boldsymbol{I}$ along the crystallographic $c$ axis, $-7/2 \leq m \leq 7/2$) [28,31]. For the $\Gamma_1$ and $\Gamma_2$ non-degenerate electronic CF states, magnetic HFS is forbidden in the first approximation. Electric quadrupole and pseudoquadrupole (magnetic dipole in the second approximation) hyperfine interactions split $\Gamma_1$ and $\Gamma_2$ singlets into four nonequidistant hyperfine sublevels and lead to nonequidistance in $\Gamma_{34}$ hyperfine manifolds [29].

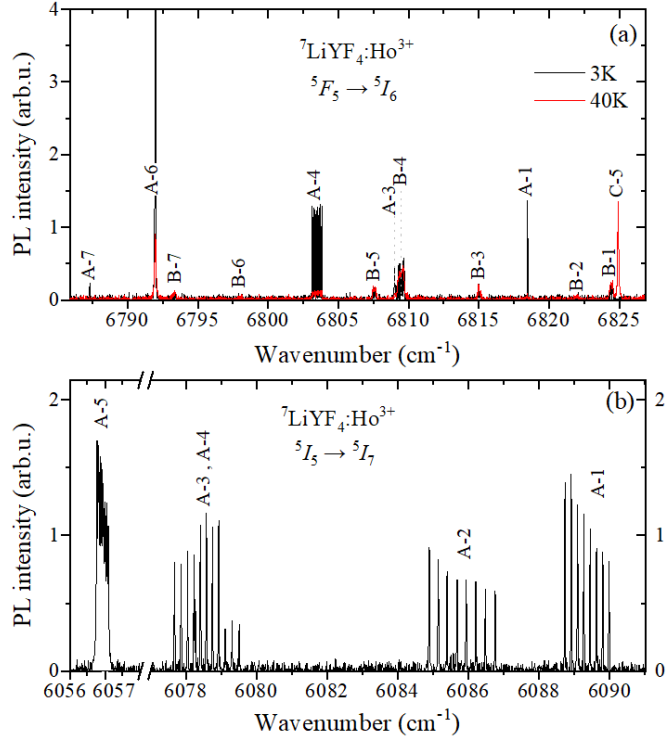

FIG. 1. Photoluminescence spectra of $^7$LiYF$_4$:Ho$^{3+}$(0.1 at.%) ($\lambda_{ex}$ = 638 nm) in the region of (a) $^5F_5 \to ^5I_6$ and (b) $^5I_5 \to ^5I_7$ transitions of Ho$^{3+}$ at the temperatures of 3 K (black spectra) and 40 K (red spectrum in (a)).

First, we choose two spectral lines corresponding to transitions from two different CF levels and demonstrate a possibility of Boltzmann ratiometric thermometry for the temperatures in the region from several to several tens Kelvin. Figure 2a shows the temperature dependence of the ratio $LIR(T)$ of the integral intensities of two lines belonging to the $^5F_5 \to ^5I_6$ luminescent multiplet and corresponding to singlet – singlet transitions, namely of the lines at 6824.95 cm$^{-1}$ [transition C($\Gamma_1$) → 5($\Gamma_2$)] and 6792 cm$^{-1}$ [transition A($\Gamma_2$) → 6($\Gamma_1$)]. This dependence follows the Boltzmann distribution of the populations of levels C and A, separated by an interval of 23.3 cm$^{-1}$ (see Fig. 6b). Figure 6c illustrates the absolute and relative sensitivities as a function of temperature. The maximum absolute sensitivity is achieved at $T_m$ = 16.8 K, with a relative sensitivity of 12% K$^{-1}$ at this temperature. The standard deviation of the experimental points in Figure 2b from the dotted line $\delta T = \pm 0.15$ K can be taken as the temperature measurement accuracy. A simple and reliable luminescence cryothermometer for the region 10 – 35 K can be implemented, with relative sensitivity decreasing from 33 to 2.7 % K$^{-1}$ in this temperature interval.

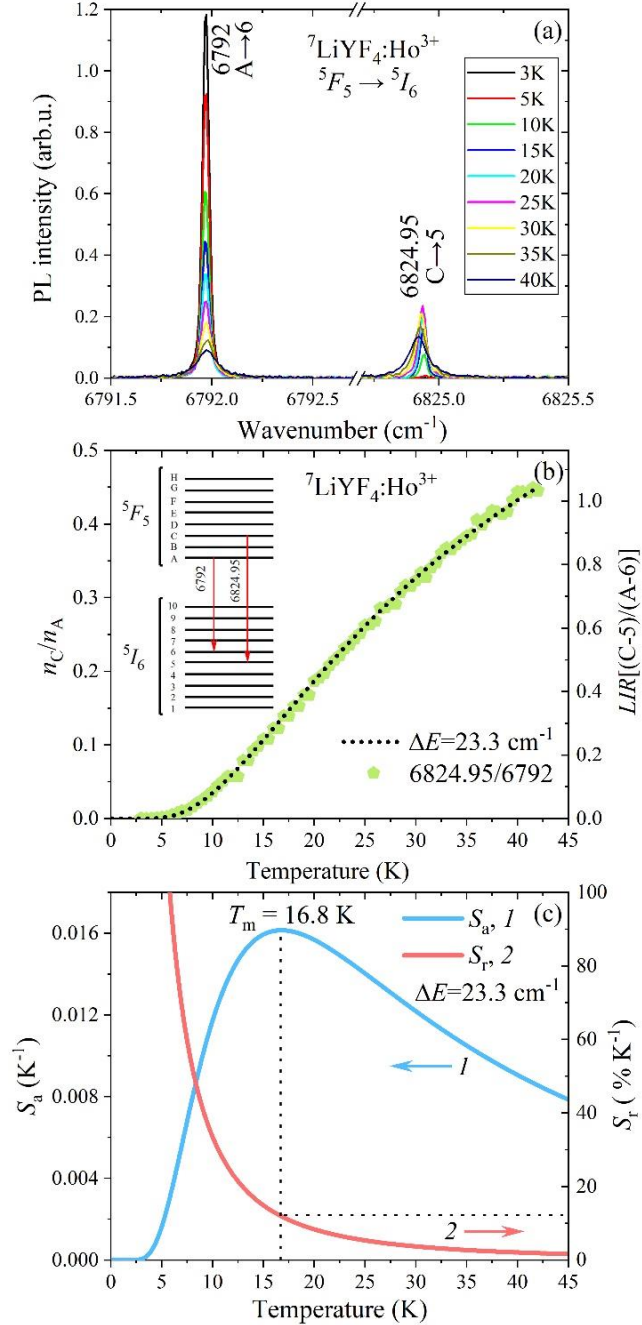

Fig. 2. (a) The luminescence lines 6792 cm$^{-1}$ [A($\Gamma_2$) → 6($\Gamma_1$)] and 6824.95 cm$^{-1}$ [C($\Gamma_1$) → 5($\Gamma_2$)] in the spectrum of $^5F_5 \rightarrow ^5I_6$ transition in $^7$LiYF$_4$:Ho$^{3+}$(0.1 at.%) at different temperatures and (b) the temperature dependences of the ratio of their integral intensities (symbols) and the Boltzmann ratio of populations $n_C/n_A$ of levels C and A separated by an interval of 23.3 cm$^{-1}$ (dotted line); (c) absolute $S_a$ (blue curve 1) and relative $S_r$ (red curve 2) sensitivities.

Next, we explore a possibility of measuring even lower temperatures using HFS. As it was already mentioned in the Introduction, HFS in the spectra of monoisotopic $^7$LiYF$_4$:Ho$^{3+}$ crystal is not complicated by an additional structure due to nonequivalent holmium centers with different $^7$Li/$^6$Li composition in the nearest surrounding of the Ho$^{3+}$ ion in a crystal with natural abundance of the lithium isotopes [32]. At very low temperatures, the luminescence spectra contain only lines starting from the lowest metastable CF levels of excited multiplets. Among these levels, only the level 11241.6 cm$^{-1}$ in the $^5I_5$ multiplet has $\Gamma_{34}$ symmetry and, thus, possesses magnetic HFS [31]. Consequently, only the luminescence lines, for which this level is the initial level of the corresponding transition, are of interest for cryothermometry with HFS. In the $^5I_5 \rightarrow ^5I_7$ luminescence manifold, several lines with resolved HFS are observed at low temperatures (see Fig. 1b). All of them have the level A ($\Gamma_{34}$, 11241.6 cm$^{-1}$) as the initial level of optical transition. This level possesses an almost equidistant HFS with extremely wide interval $\Delta_{HF} = 0.178$ cm$^{-1}$, which is favorable for cryothermometry applications. A-5 line at 6056.9 cm$^{-1}$ has the level 5 ($\Gamma_{34}$, 5184.7 cm$^{-1}$) with $\Delta_{HF} = 0.131$ cm$^{-1}$ as the final level of the transition. HFS of the A-5 line has a small interval $\Delta_{HF} = 0.047$ cm$^{-1}$, which is a difference of HFS of the levels A and 5. The lines A-4 and A-3 terminating at close singlet CF levels 4 ($\Gamma_1$, 5163.3 cm$^{-1}$) and 3 ($\Gamma_2$, 5163.8 cm$^{-1}$), respectively, superimpose and are difficult to analyze. So, we concentrate on the lines A-2 at 6085.85 cm$^{-1}$ ($\Delta_{HF} = 0.26$ cm$^{-1}$) and A-1 at 6089.3 cm$^{-1}$ ($\Delta_{HF} = 0.178$ cm$^{-1}$) with terminal levels 2 ($\Gamma_{34}$, 5155.75 cm$^{-1}$) and 1 ($\Gamma_2$, 5152.3 cm$^{-1}$) of the corresponding transitions, respectively.

It is important to note that the second-order terms in the hyperfine interaction mix wave functions of different CF levels. As this admixture depends on the energy difference between the levels, it can be different for different hyperfine components, which leads to different transition probabilities from different hyperfine components. The next Section is devoted to calculations of the intensity distributions within the selected hyperfine manifolds A-1 at 6089.3 cm$^{-1}$ and A-2 at 6085.85 cm$^{-1}$.

## IV. CALCULATION OF HYPERFINE STRUCTURE IN THE LUMINESCENCE SPECTRA OF LiYF$_4$:Ho$^{3+}$

### A. Method of calculation

Modeling of the luminescence spectra of $^7$LiYF$_4$:Ho$^{3+}$ crystal was performed using calculations of the energy levels and corresponding wave functions of the impurity ion in the crystal field. We calculated the hyperfine structure of the energy spectrum and the intensities of radiative zero-phonon transitions from the hyperfine sublevels of excited metastable states. The Hamiltonian of the Ho$^{3+}$ ion was defined in the full space of electron-nuclear states of the ground electron configuration $4f^{10}$ with the dimension $D = C_{14}^{10}(2I+1) = 8008$,

$$H = H_0 + H_{CF} + H_{HFM} + H_{HFQ} + H_{EM}. \tag{4}$$

In operator (4), the first term corresponds to the energy of a free ion [35]. The second term determines the energy of 4f electrons in the crystal field [36]. In a Cartesian coordinate system with axes x, y, z along the crystallographic axes (***a,a,c***), it can be represented as follows:

$$H_{CF} = \sum_k (B_2^0 O_{2,k}^0 + B_4^0 O_{4,k}^0 + B_6^0 O_{6,k}^0 + B_4^4 O_{4,k}^4 + B_4^{-4} O_{4,k}^{-4} + B_6^4 O_{6,k}^4 + B_6^{-4} O_{6,k}^{-4}), \tag{5}$$

where $B_p^q$ are CF parameters, $O_p^q$ are linear combinations of spherical tensor operators $C_p^q$ [37], summation is over 4f electrons. The next two terms correspond to the hyperfine magnetic dipole (HFM) and electric quadrupole (HFQ) interactions [38,39]:

$$H_{HFM} = A_{HF} \sum_k \{\mathbf{I}\mathbf{l}_k + \frac{1}{2}[O_{2,k}^0(3s_{kz}I_z - \mathbf{s}_k\mathbf{I}) + 3O_{2,k}^2(s_{kx}I_x - s_{ky}I_y) + 3O_{2,k}^{-2}(s_{kx}I_y + s_{ky}I_x)$$
$$+ 6O_{2,k}^1(s_{kx}I_z + s_{kz}I_x) + 6O_{2,k}^{-1}(s_{kz}I_y + s_{ky}I_z)]\}, \tag{6}$$

$$H_{HFQ} = \frac{e^2 Q(1-\gamma_\infty)}{4I(2I-1)} \sum_{Ls} q_s \frac{3Z_{Ls}^2 - R_{Ls}^2}{R_{Ls}^5} I_0$$
$$- \frac{e^2 Q(1-R_Q)}{4I(2I-1)} \left\langle \frac{1}{r^3} \right\rangle_{4f} \sum_k [O_{2,k}^0 I_0 + 3O_{2,k}^2 I_2 + 3O_{2,k}^{-2} I_{-2} + 6O_{2,k}^1 I_1 + 6O_{2,k}^{-1} I_{-1}]. \tag{7}$$

In Eq. (6), the constant of magnetic hyperfine interaction equals $A_{HF} = 2\mu_B \gamma_{Ho} \hbar \left\langle \frac{1}{r^3} \right\rangle_{4f}$, $\mathbf{l}_k$ and $\mathbf{s}_k$ are operators of the orbital and spin moments of the k-th electron, $\mu_B$ is the Bohr magneton, $\gamma_{Ho}/2\pi = 8.98$ MHz/T is the nuclear gyromagnetic ratio [40], and $\langle r^{-3} \rangle_{4f} = 9.7$ at. u. is an average value of the radius of a 4f electron to the power (-3) [38]. The matrices of operators $\sum_k O_{2,k}^q s_{\alpha,k}$ were constructed in the full basis of Slater determinants of the electron configuration $4f^{10}$. In Eq. (7), *e* is the elementary charge, the first term determines the interaction of the quadrupole moment of the nucleus $Q = 2.4 \cdot 10^{-28}$ m$^2$ with the gradient of the electric field of the ionic lattice in the coordinate system centered on the holmium ion under consideration. $R_{Ls}$ is the radius vector of the ion *s* in the cell *L* with the charge $eq_s$ and spherical coordinates $\theta_{Ls}$ and $\varphi_{Ls}$. The second term corresponds to the interaction of the quadrupole moment of the nucleus with the electron shell, $I_0 = 3I_z^2 - I(I+1)$, $I_2 = I_x^2 - I_y^2$, $I_{-2} = I_x I_y + I_y I_x$, $I_1 = I_x I_z + I_z I_x$, $I_{-1} = I_y I_z + I_z I_y$; $R_Q = 0.1$ and $\gamma_\infty = -80$ are Sternheimer shielding and anti-shielding factors, respectively.

The last term in Hamiltonian (4) corresponds to the interaction of 4f electrons with electromagnetic radiation,

$$H_{EM} = -\mathbf{M} \cdot \mathbf{B}(t) - \mathbf{P} \cdot \mathbf{E}(t). \tag{8}$$

Here, $\mathbf{M} = -\mu_B \sum_k (\mathbf{l}_k + 2\mathbf{s}_k)$ and $\mathbf{P} = -\sum_k e\mathbf{r}_k$ are the operators of the magnetic and electric dipole moments of the RE ion, respectively, $\mathbf{B}(t)$ and $\mathbf{E}(t)$ are the vectors of the magnetic and electric field strengths of the radiation, respectively, considered in the secondary quantization representation.

The spectral line profiles are calculated according to the quantum theory of radiation using time-dependent perturbation theory in the continuous spectrum of a system containing a holmium ion and an electromagnetic field. In the first order of perturbation theory, electric dipole transitions between energy levels of the electron $4f$ shell are forbidden by parity selection rule. However, in a crystal field of $S_4$ symmetry, this restriction is lifted by an odd, with respect to the inversion operation, component of the crystal field, which mixes the wave functions of $4f$ electrons with the wave functions of electrons virtually excited to states of opposite parity, in particular, to $5d$ and $6g$ states.

Considering the sum of the interaction operators of the RE ion with an odd crystal field and with the electric field of the radiation as a perturbation and neglecting the width of the spectrum of the mixed excited configuration $4f^9 5d$ compared to the difference in the energies of the centers of gravity of the ground ($4f^{10}$) and excited ($4f^9 5d$) configurations [41,42], we obtain the operators of the components of the effective electric dipole moment of the $4f$ electron in the following form:

$$P_x = \sum_k [b_2^1 O_{2,k}^1 + b_2^{-1} O_{2,k}^{-1} + b_4^1 O_{4,k}^1 + b_4^{-1} O_{4,k}^{-1} + b_4^3 O_{4,k}^3 + b_4^{-3} O_{4,k}^{-3}$$
$$+ b_6^1 O_{6,k}^1 + b_6^{-1} O_{6,k}^{-1} + b_6^3 O_{6,k}^3 + b_6^{-3} O_{6,k}^{-3} + b_6^5 O_{6,k}^5 + b_6^{-5} O_{6,k}^{-5}],$$

$$P_y = \sum_k [b_2^{-1} O_{2,k}^1 - b_2^1 O_{2,k}^{-1} + b_4^{-1} O_{4,k}^1 - b_4^1 O_{4,k}^{-1} - b_4^{-3} O_{4,k}^3 + b_4^3 O_{4,k}^{-3}$$
$$+ b_6^{-1} O_{6,k}^1 - b_6^1 O_{6,k}^{-1} - b_6^{-3} O_{6,k}^3 + b_6^3 O_{6,k}^{-3} + b_6^{-5} O_{6,k}^5 - b_6^5 O_{6,k}^{-5}],$$

$$P_z = \sum_k [b_2^2 O_{2,k}^2 + b_2^{-2} O_{2,k}^{-2} + b_4^2 O_{4,k}^2 + b_4^{-2} O_{4,k}^{-2}$$
$$+ b_{6,k}^2 O_6^2 + b_6^{-2} O_{6,k}^{-2} + b_6^6 O_{6,k}^6 + b_6^{-6} O_{6,k}^{-6}]. \tag{9}$$

The values of the parameters $b_p^k$, which determine the operators (9) in the space of states of the orbital momentum $l=3$, are given in Table 1.

Table 1. Parameters of the effective electric dipole moment operators $b_p^k$ (in units of $10^{-4}$ $e$ nm).

| p | k | $b_p^k$ | p | k | $b_p^k$ | p | k | $b_p^k$ |
|---|---|---|---|---|---|---|---|---|
| 2 | 1 | 6.2 | 2 | -1 | -7.7 | 2 | 2 | 1.20 |
| 4 | 1 | -8.5 | 4 | -1 | 18.1 | 2 | -2 | -0.123 |
| 6 | 1 | 2.7 | 6 | -1 | -6.2 | 4 | 2 | 3.12 |
| 4 | 3 | -20.8 | 4 | -3 | -25.6 | 4 | -2 | 0.51 |
| 6 | 3 | -6.5 | 6 | -3 | 18.9 | 6 | 2 | -2.04 |
| 6 | 5 | -28.8 | 6 | -5 | 4.6 | 6 | -2 | 13.15 |
|   |   |   |   |   |   | 6 | 6 | -1.35 |
|   |   |   |   |   |   | 6 | -6 | -0.082 |

Assuming a Lorentzian shape of the lines of individual transitions between the hyperfine components of the CF levels $\Gamma$ and $\Gamma'$ of two multiplets, the intensity distribution of spontaneous emission $\Gamma \to \Gamma'$ at temperature $T$ (the profile of the envelope of the hyperfine structure) can be described (with an accuracy of a numerical factor) by the formula

$$I(\Gamma \to \Gamma', E) = \sum_{j \in \Gamma} \sum_{k \in \Gamma'} \sum_{\alpha=x,y,z} [|<j|M_\alpha|k>|^2 + \xi|<j|P_\alpha|k|^2]n_j(T)[(E_j - E_k - E)^2 + \Delta_{\Gamma\Gamma'}^2]^{-1},$$

(10)

where $\Delta_{\Gamma\Gamma'}$ is the halfwidth of the spectral line of the electron-nuclear transition, the factor $\xi = (n^2+2)/9n^2$ includes a correction for the Lorentz field ($n = 1.44$ is the refractive index).

**B. Results of calculations of spectral line profiles**

The energies of the electron-nuclear states corresponding to the five lower electronic multiplets $^4I_J$ and $^5F_5$ of the Ho$^{3+}$ ion were obtained by numerical diagonalization of the Hamilton operator (4) projected onto the space of 608 electron-nuclear wave functions constructed on the basis of 76 electronic functions of the considered CF levels. The energies of the CF levels of the electronic multiplets were corrected in accordance with the measurement data [33]. The applied calculation method ensured that the hybridization of the wave functions of the hyperfine sublevels of the electronic multiplets, caused by the hyperfine interaction, was taken into account.

The calculations used the magnetic hyperfine interaction constant $A_{HF}=0.03663$ cm$^{-1}$ and the optimized parameters $B_{HF}=1.5886 \cdot 10^{-4}$ cm$^{-1}$, $D_{HF}=2.892 \cdot 10^{-3}$ cm$^{-1}$ of the electric quadrupole interaction operator

$$H_{HFQ} = -B_{HF}I_0 - D_{HF} \sum_k (O_{2,k}^0 I_0 + 3O_{2,k}^2 I_2 + 3O_{2,k}^{-2} I_{-2} + 6O_{2,k}^1 I_1 + 6O_{2,k}^{-1} I_{-1}).$$

(11)

We have calculated HFS profiles of the transitions from the lower doublet $\Gamma_{34}$ in the $^5I_5$ multiplet to the lower singlet $\Gamma_2$ and the closest doublet $\Gamma_{34}$ with an energy difference of 3.5 cm$^{-1}$ in the $^5I_7$ multiplet at temperatures of 3, 5, 10, 15, 20, and 25 K. The half-width of individual transitions between the hyperfine sublevels of the doublet $\Gamma_{34}$ ($^5I_5$) and the hyperfine sublevels of the lower CF components of the $^5I_7$ multiplet in the distribution (10) was taken to be temperature-independent in this temperature region and equal to $\Delta_{\Gamma\Gamma'} = 6 \cdot 10^{-3}$ cm$^{-1}$. The populations of the sublevels $n_j(T)$ of the metastable doublet $\Gamma_{34}$ ($^5I_5$) in Eq. (10) were specified by the Boltzmann distribution at the corresponding temperature with energies measured from the lower sublevel of this doublet. Figure 3 compares the calculated (black dashed lines) and measured (solid red lines) hyperfine structure profiles at nominal temperature of 3 K.

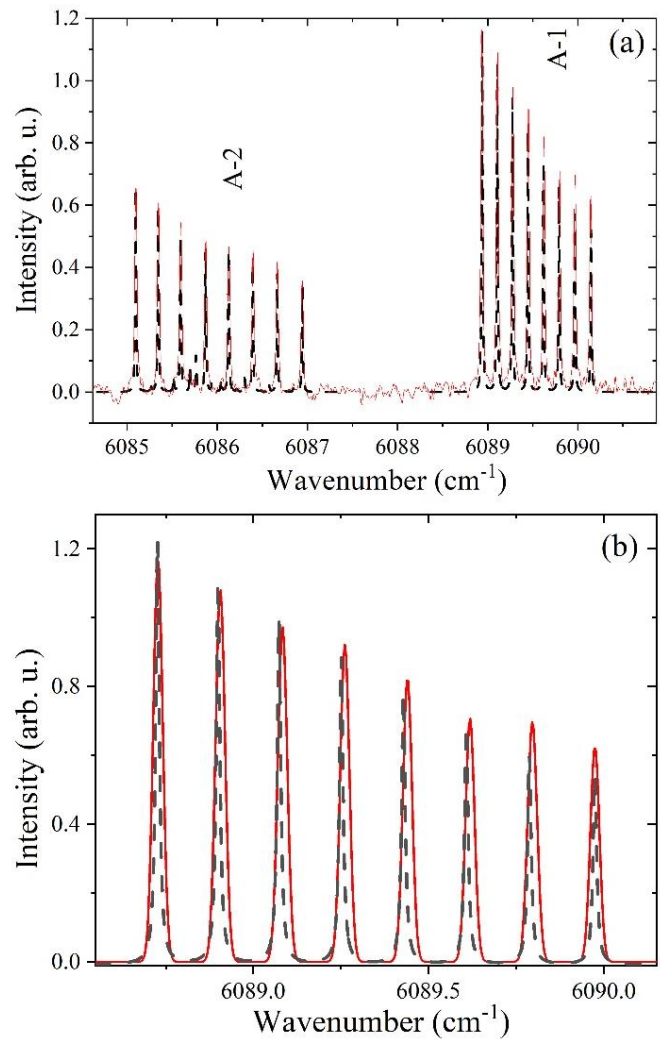

Fig. 3. Experimental (red, $\lambda_{ex}$ = 638 nm, resolution 0.02 cm$^{-1}$) and calculated (black dashed) PL spectra of $^7$LiYF$_4$:Ho$^{3+}$(0.1 at.%) at $T$ = 3 K in the region of the lines (a) A-2 [$^5I_5$ $\Gamma_{34}$, 11241.6 cm$^{-1}$ → $^5I_7$ $\Gamma_{34}$, 5155.75 cm$^{-1}$] and A-1 [$^5I_5$ $\Gamma_{34}$, 11241.6 cm$^{-1}$ → $^5I_7$ $\Gamma_2$, 5152.3 cm$^{-1}$] and (b) A-1 on an enlarged scale.

## V. TEMPERATURE SENSING BASED ON THE INTENSITY DISTRIBUTION OF HYPERFINE COMPONENTS

The analysis of the calculated spectra has shown that the temperature-dependent intensities $I(v_j)$ of hyperfine components within HFS of the line A-1 near 6089 cm$^{-1}$ could be well approximated by the following empirical relation:

$$I(v_j, T) = I_1 \exp[-b(v_j - v_1)], \quad b = a + 1/kT, \quad j = 1, 2, \ldots 8, \tag{12}$$

where $I_1$ is the intensity of the first (lowest-frequency) hyperfine component, $v_j$ is the frequency of the j-th hyperfine component, and $a = 0.18$ cm accounts for the admixture of the wave functions of the nearest CF levels, caused by the hyperfine interaction. Eq. (12) leads to a linear dependence at the logarithmic scale:

$$\ln I(v_j, T) = \ln I_1 - b(v_j - v_1), \quad j = 1, 2, \ldots 8. \tag{13}$$

In Fig. 4, such approximation is shown for two temperatures.

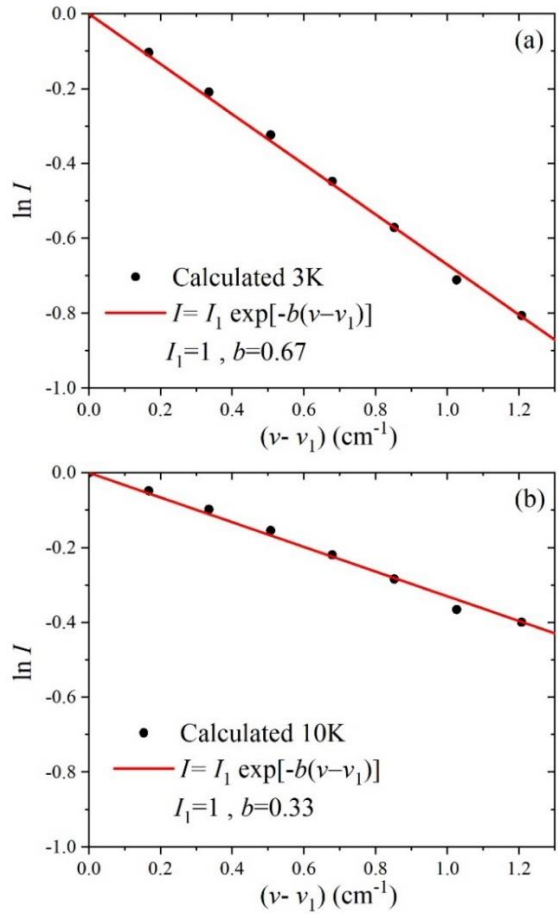

Fig. 4. Approximation by Eq. (13) of the calculated intensities (black dots) of hyperfine components within HFS of the line A-1 near 6089 cm$^{-1}$ [$^5I_5$ $\Gamma_{34}$, 11241.6 cm$^{-1}$ → $^5I_7$ $\Gamma_2$, 5152.3 cm$^{-1}$] in PL spectra of $^7$LiYF$_4$:Ho$^{3+}$ at the temperatures (a) 3 K and (b) 10K.

No simple relation could be found to approximate the intensity distribution within the HFS of the line A-2 near 6086 cm$^{-1}$. So, we choose the HFS of the line A-1 near 6089 cm$^{-1}$ for low-temperature measurements. We propose to find the value of $b$ from the approximation of the measured intensities of hyperfine components by Eq. (13) using the least squares method and to calculate the value of temperature as $T = 1/(b-a)k$. In this method, the absolute thermal sensitivity

$$S_a(T) = |\frac{d[b(T)]}{dT}| = \frac{1}{kT^2} \tag{14}$$

grows rapidly at $T$ approaching zero. The relative sensitivity behaves as $1/T$ at $T \to 0$ and as $1/akT^2$ at $T \to \infty$.

Figure 5 compares the temperature dependences of the relative intensities of the highest and the lowest in frequency components of HFS of the line A-1 near 6089 cm$^{-1}$ (yellow symbols) and the relative populations of the corresponding hyperfine levels, assuming the Boltzmann distribution (black solid line). This comparison confirms the fulfilment of the Boltzmann distribution for the populations of hyperfine sublevels, within the precision of measurements.

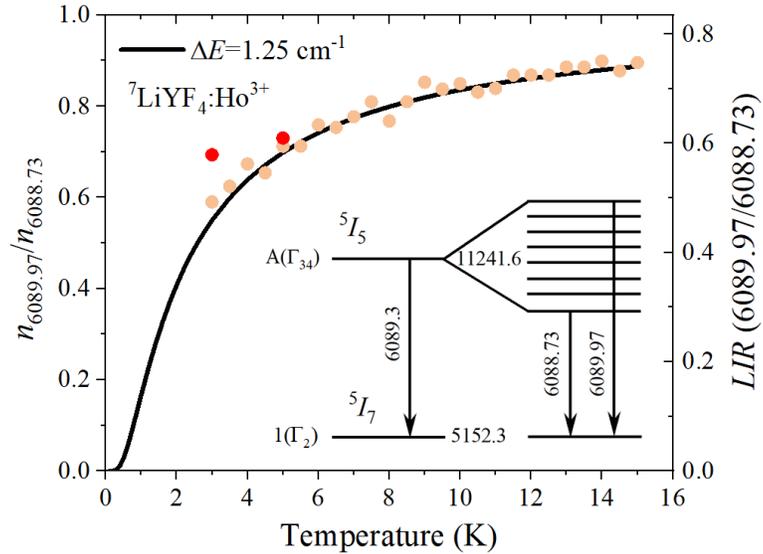

Fig. 5. Temperature dependence of the relative integral intensities of the highest and the lowest in frequency components of HFS of the PL line A-1 near 6089 cm$^{-1}$ (yellow symbols). Black solid line represents the relative populations of the corresponding hyperfine levels, assuming the Boltzmann distribution. Red dots are from a higher resolution (0.01 cm$^{-1}$) spectra obtained without taking precautions to reduce heating.

Next, we specify experimentally the theoretical value of the quantity $a$ responsible for different radiative transition probabilities from different hyperfine components of the initial level $^5I_5$ $\Gamma_{34}$,

11241.6 cm$^{-1}$ of the considered transition. For that, we analyze the spectra at sufficiently high temperatures ($T$ = 8, 9, 10, 11, 12, 13, and 14 K) to exclude a possible heating of the sample by the laser radiation used to excite PL. The hyperfine components in the spectra were approximated by Gaussians with equal widths for all components. The maximums of these Gaussians were taken as experimental values $I(v_j, T)$. Applying Eq. (13) and using the nominal values of temperatures, we obtained the value $a$ = (0.169 ± 0.001) cm.

Figure 6 shows the experimental values of the intensities of hyperfine components in the HFS of the A-1 line in the PL spectrum of $^7$LiYF$_4$:Ho$^{3+}$(0.1 at.%) at the nominal temperature of 3 K and the approximation of these eight experimental points by the straight line (13) using the least squares method. The obtained value of the slope of the straight line $b$ = 0.56 ± 0.02 results in the temperature $T$ = 3.7 ± 0.2 K.

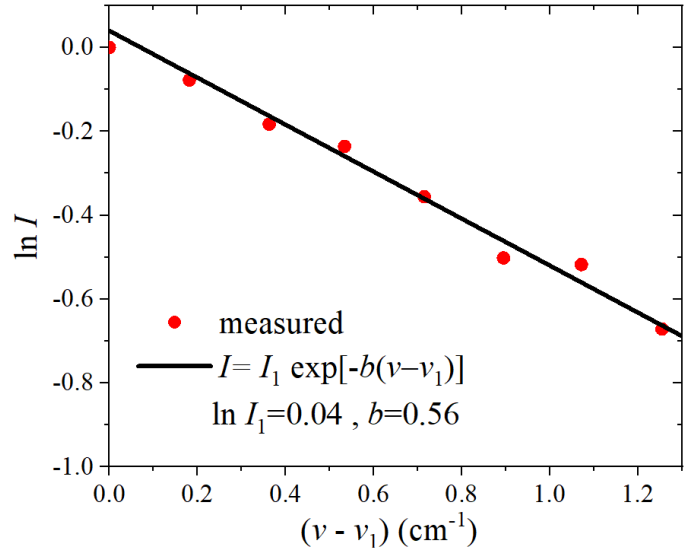

Fig. 6. Approximation by Eq. (13) of the measured intensities (red dots) of hyperfine components within HFS of the line A-1 near 6089 cm$^{-1}$ [$^5I_5$ $\Gamma_{34}$, 11241.6 cm$^{-1}$ → $^5I_7$ $\Gamma_2$, 5152.3 cm$^{-1}$] in PL spectrum of $^7$LiYF$_4$:Ho$^{3+}$(0.1 at.%) ($\lambda_{ex}$ = 638 nm).

The difference between thus determined temperature and the nominal one is, most likely, caused by a heating of the sample by the incident laser radiation. Red dots in Fig. 5 refer to higher resolution spectra obtained without taking precautions to reduce heating. It is evident that to measure temperatures below of about 5 K, one should carefully select the wavelength of the exciting radiation to avoid heating. As for the precision of the temperature measurement, it can be evaluated using Eq. (14) as $\Delta T_1 / \Delta T_2 = (T_1 / T_2)^2$, provided $\Delta b$ does not depend on temperature. For example, while $\Delta T$ = ± 0.2 K for $T$ = 3 K, it is reasonable to expect $\Delta T$ = ±0.02 K for $T$ = 1 K.

## VII. CONCLUSIONS

We have proposed to use the measured intensity distribution within a hyperfine manifold in the luminescence spectra of a holmium-doped crystal for very low temperature measurement. To realize this strategy, it is necessary (i) to use a metastable CF level with well-resolved HFS as an initial level of the luminescent transition; (ii) to know the distribution of optical transition probabilities within a selected hyperfine manifold, which depends on intermixing of wave functions of different electronic CF levels by the hyperfine interaction.

A monoisotopic $^7$LiYF$_4$:Ho$^{3+}$ crystal was chosen. HFS of some luminescence lines corresponding to optical transitions involving doubly degenerate electronic $\Gamma_{34}$ crystal-field levels in the PL spectra of $^7$LiYF$_4$:Ho$^{3+}$ at low temperatures consists of eight well resolved narrow components and is not complicated by additional lines appearing due to isotopic disorder in the lithium sublattice of a crystal with natural abundance of lithium isotopes. While the energy structure of the spectra was comprehensively studied previously, both experimentally and theoretically, the intensity distributions within hyperfine manifolds was not studied before. Here, we calculated the hyperfine structure of the energy spectrum and the intensities of radiative zero-phonon transitions from the hyperfine sublevels of excited metastable states to create a basis for developing a method of cryothermometry based on HFS intensity measurements.

The infrared line at 6089.3 cm$^{-1}$ of the $^5I_5 \rightarrow {}^5I_7$ holmium transition falling into the transparency window of optical fibers (the S telecom band) was selected. The reasons for this selection are the following: (i) the initial metastable CF level of the corresponding optical transition $^5I_5$ $\Gamma_{34}$, 11241.6 cm$^{-1}$ has an almost equidistant HFS with extremely wide interval $\Delta_{HF} = 0.178$ cm$^{-1}$, which is favorable for cryothermometry applications; (ii) among the PL lines starting from this level, the line 6089.3 cm$^{-1}$ has the distribution of transition probabilities within HFS, which can be easily taken into account. The calculated intensity distribution $I(\nu_j)$, j=1, 2, …8, between the hyperfine components of the line 6089.3 cm$^{-1}$ is well approximated by an exponential function whose index is the sum of a temperature-independent term responsible for the wave function intermixing and a term responsible for the temperature-independent distribution of populations of hyperfine sublevels. This half-empirical formula was verified experimentally.

To determine the temperature $T$, we perform the least-square fitting of the eight points $y_j = \ln I(\nu_j)$ (corresponding to experimentally measured intensities of the hyperfine components) by a straight line $y(\nu,T) = A - b\nu, \quad b = a + 1/kT,$ where $1/kT$ comes from the Boltzmann factor but $a$ accounts for intensity variations due to mixing of wave functions of different crystal-field levels by the hyperfine interaction. The temperature is found from the slope $b$ of the obtained straight line. The described procedure when applied to the line 6089.3 cm$^{-1}$ in the PL spectrum of $^7$LiYF$_4$:Ho$^{3+}$ at a nominal temperature of 3K resulted in the value of $T = 3.7 \pm 0.2$ K, which exceeds the nominal temperature due to heating of the sample by laser radiation. Further work is necessary to eliminate this heating. We plan to use a narrow excitation line from a Ti-sapphire tunable laser. As the absolute thermal sensitivity of the method $S_a(T) = |db/dT| = 1/kT^2$ grows with lowering the temperature, one could expect $\Delta T = \pm 0.02$ K for $T = 1$ K, supposing the accuracy of determining $b$ was same as in the case of $T = 3$ K.

**Acknowledgements**

The authors express their gratitude to V. J. Egorov and M. A. Petrova for growing crystals, as well as to K. N. Boldyrev and N. Yu. Boldyrev for assistance in carrying out the measurements.

The work was supported by the Russian Science Foundation under Grant No 23-12-00047.

**Supplemental Material for**

**High-precision luminescent cryothermometry strategy by using hyperfine structure**

Marina N. Popova[1]*, Mosab Diab[1,2], and Boris Z. Malkin[3]

[1] *Institute of Spectroscopy, Russian Academy of Sciences, Troitsk, Moscow 108840, Russia*

[2] *Moscow Institute of Physics and Technology (National Research University), Dolgoprudnyi, 141700, Russia*

[3] *Kazan Federal University, Kazan, 420008, Russia*

**1**. Identification of lines observed in the $^5F_5 \rightarrow {}^5I_6$ and $^5I_5 \rightarrow {}^5I_7$ spectral multiplets (**Tables S1 and S2**)

**2**. Selection rules (**Table S3**)

# 1. Identification of lines observed in the $^5F_5 \to {}^5I_6$ and $^5I_5 \to {}^5I_7$ spectral multiplets

Energies (in cm$^{-1}$) and symmetries (IRs) of crystal-field (CF) levels of Ho$^{3+}$ in LiYF$_4$:Ho$^{3+}$ and intervals (cm$^{-1}$) of magnetic dipole HFS of non-Kramers doublets $\Gamma_{34}$ are taken from Refs. [S1-S3]. Frequencies of transitions allowed only as magnetic-dipole ones are highlighted in grey in Tables S1 and S2.

## Table S1. $^5F_5 \to {}^5I_6$

| $^5I_6$ \ $^5F_5$ | | A<br>15489.4<br>($\Gamma_2$) | B<br>15495.4<br>($\Gamma_{34}$)<br>0.034 | C<br>15512.7<br>($\Gamma_1$) | D<br>15558.7<br>($\Gamma_1$) | E<br>15623.1<br>($\Gamma_{34}$) | F<br>15632.1<br>($\Gamma_2$) | G<br>15639.7<br>($\Gamma_1$) | H<br>15667.2<br>($\Gamma_{34}$) |
|---|---|---|---|---|---|---|---|---|---|
| 10 | 8796.6 ($\Gamma_2$) | 6692.8 | 6698.8 | 6716.1 | 6762.1 | 6826.5 | 6835.5 | 6843.1 | 6870.6 |
| 9 | 8783.7 ($\Gamma_{34}$) 0.06 | 6705.7 | 6711.7 | 6729 | 6775 | 6839.4 | 6848.4 | 6856 | 6883.5 |
| 8 | 8769.0 ($\Gamma_1$) | 6720.4 | 6726.4 | 6743.7 | 6789.7 | 6854.1 | 6863.1 | 6870.7 | 6898.2 |
| 7 | 8702.03 ($\Gamma_2$) | 6787.37 | 6793.37 | 6810.67 | 6856.67 | 6921.07 | 6930.07 | 6937.67 | 6965.17 |
| 6 | 8697.4 ($\Gamma_1$) | 6792 | 6798 | 6815.3 | 6861.3 | 6925.7 | 6934.7 | 6942.3 | 6969.8 |
| 5 | 8687.75 ($\Gamma_2$) | 6801.65 | 6807.65 | 6824.95 | 6870.95 | 6935.35 | 6944.35 | 6951.95 | 6979.45 |
| 4 | 8685.9 ($\Gamma_{34}$) 0.095 | 6803.5 | 6809.5 | 6826.8 | 6872.8 | 6937.2 | 6946.2 | 6953.8 | 6981.3 |
| 3 | 8680.3 ($\Gamma_{34}$) 0.015 | 6809.1 | 6815.1 | 6832.4 | 6878.4 | 6942.8 | 6951.8 | 6959.4 | 6986.9 |
| 2 | 8673.4 ($\Gamma_1$) | 6816 | 6822 | 6839.3 | 6885.3 | 6949.7 | 6958.7 | 6966.3 | 6993.8 |
| 1 | 8670.9 ($\Gamma_2$) | 6818.5 | 6824.5 | 6841.8 | 6887.8 | 6952.2 | 6961.2 | 6968.8 | 6996.3 |

**Table S2.** $^5I_5 \rightarrow {}^5I_7$

| $^5I_7$ \ $^5I_5$ | | A<br>11241.6<br>($\Gamma_{34}$)<br>0.178 | B<br>11247.2<br>($\Gamma_1$) | C<br>11249.9<br>($\Gamma_{34}$)<br>0.069 | D<br>11254.0<br>($\Gamma_2$) | E<br>11255.6<br>($\Gamma_2$) | F<br>11301.0<br>($\Gamma_1$) | G<br>11330.0<br>($\Gamma_{34}$)<br>0.042 | H<br>11335.9<br>($\Gamma_2$) |
|---|---|---|---|---|---|---|---|---|---|
| 11 | 5293.1 ($\Gamma_1$) | 5948.5 | 5954.1 | 5956.8 | 5960.9 | 5962.5 | 6007.9 | 6036.9 | 6042.8 |
| 10 | 5292.9 ($\Gamma_{34}$) | 5948.7 | 5954.3 | 5957 | 5961.1 | 5962.7 | 6008.1 | 6037.1 | 6043 |
| 9 | 5291.0 ($\Gamma_2$) | 5950.6 | 5956.2 | 5958.9 | 5963 | 5964.6 | 6010 | 6039 | 6044.9 |
| 8 | 5232.6 ($\Gamma_2$) | 6009 | 6014.6 | 6017.3 | 6021.4 | 6023 | 6068.4 | 6097.4 | 6103.3 |
| 7 | 5227.8 ($\Gamma_{34}$)<br>0.08 | 6013.8 | 6019.4 | 6022.1 | 6026.2 | 6027.8 | 6073.2 | 6102.2 | 6108.1 |
| 6 | 5206.1 ($\Gamma_1$) | 6035.5 | 6041.1 | 6043.8 | 6047.9 | 6049.5 | 6094.9 | 6123.9 | 6129.8 |
| 5 | 5184.7 ($\Gamma_{34}$)<br>0.131 | 6056.9 | 6062.5 | 6065.2 | 6069.3 | 6070.9 | 6116.3 | 6145.3 | 6151.2 |
| 4 | 5163.3 ($\Gamma_1$) | 6078.3 | 6083.9 | 6086.6 | 6090.7 | 6092.3 | 6137.7 | 6166.7 | 6172.6 |
| 3 | 5163.8 ($\Gamma_2$) | 6078.8 | 6084.4 | 6087.1 | 6091.2 | 6092.8 | 6138.2 | 6167.2 | 6173.1 |
| 2 | 5155.75 ($\Gamma_{34}$)<br>0.082 | 6085.85 | 6091.45 | 6094.15 | 6098.25 | 6099.85 | 6145.25 | 6174.25 | 6180.15 |
| 1 | 5152.3 ($\Gamma_2$) | 6089.3 | 6094.9 | 6097.6 | 6101.7 | 6103.3 | 6148.7 | 6177.7 | 6183.6 |

## 2. Selection rules

**Table S3.** Selection rules. $d_k$ ($m_k$), k = x, y, z, denote the allowed components of the ED (MD) transitions. For convenience, also polarizations are indicated, e.g. $\sigma_e \pi_m$ means that the transition is ED allowed in the $\sigma$ polarization ($\mathbf{k} \perp c$, $\mathbf{E} \perp c$) and MD allowed in the $\pi$ polarization ($\mathbf{k} \perp c$, $\mathbf{E} \| c$).

| $S_4$ | $\Gamma_1$ | $\Gamma_2$ | $\Gamma_3$ | $\Gamma_4$ |
|---|---|---|---|---|
| $\Gamma_1$ | $m_z$<br>$\sigma_m$ | $d_z$<br>$\pi_e$ | $d_{x-iy}, m_{x+i,y}$<br>$\alpha_e \sigma_e \; \alpha_m \pi_m$ | $d_{x+iy}, m_{x-i,y}$ |
| $\Gamma_2$ | | $m_z$<br>$\sigma_m$ | $d_{x+iy}, m_{x-i,y}$<br>$\alpha_e \sigma_e \; \alpha_m \pi_m$ | $d_{x-iy}, m_{x+i,y}$ |
| $\Gamma_3$ | | | $m_z$<br>$\sigma_m$ | $d_z$<br>$\pi_e$ |
| $\Gamma_4$ | | | | $m_z$<br>$\sigma_m$ |